\RequirePackage{rotating}
\documentclass{article}

\usepackage{arxiv}

\usepackage[utf8]{inputenc} % allow utf-8 input
\usepackage[T1]{fontenc}    % use 8-bit T1 fonts
\usepackage{hyperref}       % hyperlinks
\usepackage{url}            % simple URL typesetting
\usepackage{booktabs}       % professional-quality tables
\usepackage{amsfonts}       % blackboard math symbols
\usepackage{nicefrac}       % compact symbols for 1/2, etc.
\usepackage{microtype}      % microtypography
\usepackage{lipsum}

\usepackage{array}
\usepackage{acronym}
\usepackage{tabularx}
\usepackage[inline]{enumitem}

\acrodef{AI}{artificial intelligence}
\acrodef{ALVP}[ALV project]{Autonomous Land driven Vehicle project}
\acrodef{ANN}{artificial neural network}
\acrodef{ASIL}{Automotive Safety Integrity Level}
\acrodef{DARPA}{Defense Advanced Research Projects Agency}
\acrodef{GAN}{Generative adversarial network}
\acrodef{ISO}{International Organization for Standardization}
\acrodef{SAE}{Society of Automotive Engineers}
\acrodef{V2I}{vehicle-to-infrastructure}
\acrodef{V2V}{Vehicle-to-vehicle}

\title{Validation Frameworks for Self-Driving Vehicles: A Survey}

\author{Francesco Concas, Jukka K. Nurminen, Tommi Mikkonen, and Sasu Tarkoma}

\author{
  Francesco Concas \\
  Department of Computer Science \\
  University of Helsinki \\
  PL 64, Helsinki, Finland \\
  \texttt{francesco.concas@helsinki.fi} \\
   \And
  Jukka K. Nurminen \\
  Department of Computer Science \\
  University of Helsinki \\
  PL 68, Helsinki, Finland \\
  \texttt{jukka.k.nurminen@helsinki.fi} \\
   \And
  Tommi Mikkonen \\
  Department of Computer Science \\
  University of Helsinki \\
  PL 68, Helsinki, Finland \\
  \texttt{tommi.mikkonen@helsinki.fi} \\
   \And
  Sasu Tarkoma \\
  Department of Computer Science \\
  University of Helsinki \\
  PL 68, Helsinki, Finland \\
  \texttt{sasu.tarkoma@helsinki.fi} \\
}

\begin{document}
\maketitle

\begin{abstract}
As a part of the digital transformation, we interact with more and more intelligent gadgets.
Today, these gadgets are often mobile devices, but in the advent of smart cities, more and more infrastructure---such as traffic and buildings---in our surroundings becomes intelligent.
The intelligence, however, does not emerge by itself.
Instead, we need both design techniques to create intelligent systems, as well as approaches to validate their correct behavior.
An example of intelligent systems that could benefit smart cities are self-driving vehicles.
Self-driving vehicles are continuously becoming both commercially available and common on roads.
Accidents involving self-driving vehicles, however, have raised concerns about their reliability.
Due to these concerns, the safety of self-driving vehicles should be thoroughly tested before they can be released into traffic.
To ensure that self-driving vehicles encounter all possible scenarios, several millions of hours of testing must be carried out;
therefore, testing self-driving vehicles in the real world is impractical.
There is also the issue that testing self-driving vehicles directly in the traffic poses a potential safety hazard to human drivers.
To tackle this challenge, validation frameworks for testing self-driving vehicles in simulated scenarios are being developed by academia and industry.
In this chapter, we briefly introduce self-driving vehicles and give an overview of validation frameworks for testing them in a simulated environment.
We conclude by discussing what an ideal validation framework at the state of the art should be and what could benefit validation frameworks for self-driving vehicles in the future.


\end{abstract}

% keywords can be removed

\section{Introduction to Self-Driving Vehicles}\label{sec:introduction}

An intelligent infrastructure that helps humans in their daily activities is a key feature of smart cities.
Traffic is one of the biggest challenges of cities; smart cities could greatly benefit from a traffic infrastructure that can intelligently and autonomously manage it.
Self-driving vehicles, or autonomous vehicles, are vehicles capable of driving autonomously with little or no human input.
A network of self-driving vehicles able to communicate with each other have the potential to tackle this challenge and greatly improve the mobility in smart cities~\cite{Nikitas2017}.
In recent years, self-driving vehicles have become more and more commercially easily available.
A large number of self-driving vehicles now populate American and European roads, and projections show that their number will likely keep increasing~\cite{IHS2018}.
As of 2019, twenty-nine states in the United States of America have enacted legislation related to autonomous vehicles~\cite{ncsllaws}.

Autonomous driving is achieved by using a combination of sensing units that sense the surrounding environment, and an advanced control system that interprets the output from the sensing units and drives the vehicle.
The \ac{SAE} defined a standard for the levels of automation of self-driving vehicles, shown in Table~\ref{tab:SAE_EJ3016}.
In this chapter, we focus on self-driving vehicles belonging to levels of automation 4 and 5, where the system takes care of the driving commands, the monitoring of the environment, and has also full responsibility during emergencies.
In other words, we focus on vehicles for which the driver is left completely out of the loop.

\begin{sidewaystable*}
    \centering
    \caption{Levels of driving automation, as defined by the SAE EJ3016 standard~\cite{SAEJ3016B}.}
    \label{tab:SAE_EJ3016}
    \begin{tabular}{|>{\centering\arraybackslash}m{2em}|>{\centering\arraybackslash}m{5em}|>{\raggedright\arraybackslash}m{23em}|>{\centering\arraybackslash}m{7em}|>{\centering\arraybackslash}m{7em}|>{\centering\arraybackslash}m{7em}|>{\centering\arraybackslash}m{7em}|}
        \hline
            SAE level &
            Name &
            Narrative Definition &
            Execution of Steering and Acceleration/Deceleration & 
            Monitoring of Driving Environment &
            Fallback Performance of Dynamic Driving Task &
            System Capability (Driving Modes) \\
        \hline \hline
        \multicolumn{3}{|p{30em}|}{Human driver monitors the driving environment} & \multicolumn{4}{c|}{} \\
        \hline
            0 &
            No Automation &
            the full-time performance by the human driver of all aspects of the dynamic driving task, even when enhanced by warning or intervention systems &
            Human driver &
            Human driver &
            Human driver &
            n/a \\
        \hline
            1 &
            Driver Assistance &
            the driving mode-specific execution by a driver assistance system of either steering or acceleration/deceleration using information about the driving environment and with the expectation that the human driver perform all remaining aspects of the dynamic driving task &
            Human driver and system &
            Human driver &
            Human driver &
            Some driving modes \\
        \hline
            2 &
            Partial Automation &
            the driving mode-specific execution by one or more driver assistance systems of both steering and acceleration/deceleration using information about the driving environment and with the expectation that the human driver perform all remaining aspects of the dynamic driving task &
            System &
            Human driver &
            Human driver &
            Some driving modes \\
        \hline \hline
        \multicolumn{3}{|p{30em}|}{Automated driving system monitors the driving environment} & \multicolumn{4}{c|}{} \\
        \hline
            3 &
            Conditional Automation &
            the driving mode-specific performance by an automated driving system of all aspects of the dynamic driving task with the expectation that the human driver will respond appropriately to a request to intervene &
            System &
            System &
            Human driver &
            Some driving modes \\
        \hline
            4 &
            High Automation &
            the driving mode-specific performance by an automated driving system of all aspects of the dynamic driving task, even if a human driver does not respond appropriately to a request to intervene &
            System &
            System &
            System &
            Some driving modes \\
        \hline
            5 &
            Full Automation &
            the full-time performance by an automated driving system of all aspects of the dynamic driving task under all roadway and environmental conditions that can be managed by a human driver &
            System &
            System &
            System &
            All driving modes \\
        \hline
    \end{tabular}
\end{sidewaystable*}

The software that governs self-driving vehicles is, by the nature of the problem, very complex.
The software has to elaborate extremely complex input coming from numerous sensors present in the vehicle, including cameras, LIDAR, and RADAR.
Furthermore, the software has to drive the vehicle through lanes, traffic, intersections, and different weather conditions.
For these reasons, validating self-driving vehicles is also a very complex task.
Testing self-driving vehicles on public roads is unpractical.
Firstly, it would take millions of hours of driving for the vehicle to encounter every possible scenario that can happen on a public road.
Secondly, if an unforeseen fault occurs while the vehicle is driving on a public road, the vehicle poses a hazard to human drivers in the traffic.

Validation frameworks are being developed to tackle the challenge of testing self-driving vehicles.
They work by simulating self-driving vehicles, including all of their sensors, on simulated scenarios that a self-driving vehicle can encounter in a public road.
This way, potentially hazardous situations can be tested in a simulated environment so that the behavior of a self-driving vehicle can be tested without introducing a hazard on public roads.

\subsection{Components of Self-Driving Vehicles}

\begin{figure}[t]
    \centering
    \includegraphics[width=.5\linewidth]{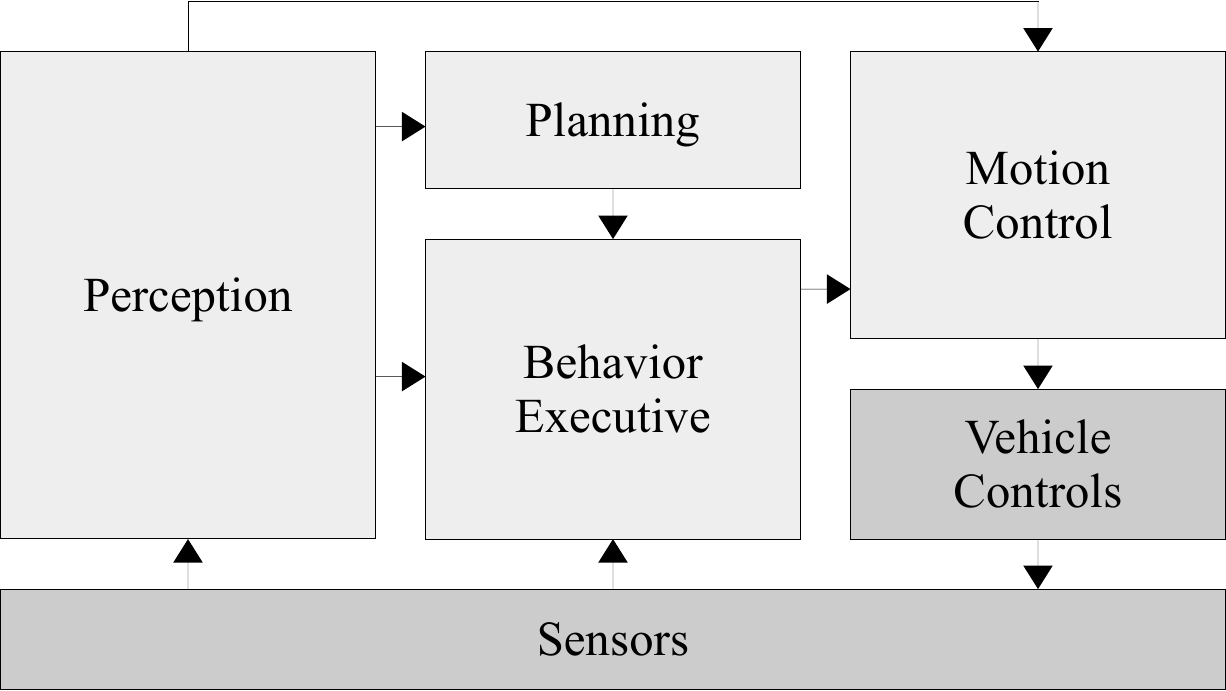}
    \caption{Generic architecture of self driving vehicles.}
    \label{fig:architecture}
\end{figure}

A self-driving vehicle, in addition to the mechanical components present in all vehicles, includes sensors, processing units, and the self-driving unit.
A self-driving vehicle perceives its surrounding environment through multiple sensors, such as cameras, LIDAR, RADAR, and GPS components~\cite{Jo2015}.
The outputs of the sensors are processed by on-board processing units, such as CPUs, GPUs, and FPGAs~\cite{Jo2015}.
To communicate with the processing units, the sensors are connected to them via multiple buses, which are in turn connected through gateways~\cite{Zheng2016a}.
The self-driving unit is the software that, using the information given by the sensors, controls the mechanical components to drive the vehicle.
Usually, the self-driving unit of a self-driving vehicle consists of four main components:
\begin{enumerate*}[label=\alph*)]
  \item a perception unit;
  \item a planning unit;
  \item a behavioral executive unit;
  \item and a motion control unit~\cite{Zheng2016a}.
\end{enumerate*}
A generic architecture of self-driving vehicles is shown in Figure~\ref{fig:architecture}.

\textbf{Perception}.
This unit receives the outputs from the sensors and produces information regarding the road shape, the position of the vehicle, road signals, and obstacles such as other vehicles, pedestrians, and animals. The output of the perception unit is then sent to every other component of the self-driving unit.
This is a very complex computer vision problem, which in turn requires a very complex machine learning algorithm.
Deep learning has shown promising results in computer vision; a related survey can be found in Voulodimos et al.~\cite{Voulodimos2018}.

\textbf{Planning}.
This unit receives information from the perception unit and plans the route, trying to minimize the time considering variables such as route length, speed limits, and traffic.
An example of an algorithm that takes into account all these variables to plan an optimal route is the Google Maps route planner~\cite{Geisberger2014}.
The output of this unit is sent to the behavioral executive unit.

\textbf{Behavioral executive}.
This unit receives information from the perception unit, the planning unit, and sensors, and is in charge of making driving decisions such as in which lane to drive, how to behave in intersections, how to move through traffic, and where to park and how to behave in parking lots.
The output of this unit is sent to the motion control unit.
To make these decisions, the behavioral executive unit needs to take into account also traffic laws.
In the robotics field, this is known as the roadmap planning problem, in which statistical methods have shown promising results~\cite{Geraerts2004}.

\textbf{Motion control}.
This unit receives information from the perception and behavioral executive units and physically executes the commands by controlling the mechanical parts of the vehicle.
This unit needs to take into account the comfort of the passengers, to reduce the chance of motion sickness and injuries caused by sudden accelerations.
Internal sensors located in the mechanical parts are used to send feedback to the units that receive inputs from the sensors, namely the perception and the behavioral executive units.

\subsection{Software-related issues}

The software needed to govern a self-driving vehicle is very complex, due to the nature of the autonomous driving problem.
Let us discuss some common issues of this kind of software.

\textbf{Black-box logic}.
Self-driving cars rely on complex machine learning models, as discussed in Section~\ref{sec:introduction}.
These models, usually belonging to the class of \acp{ANN}, rely on very complex connections in the hidden layers, which cannot be manually set but are instead ``trained'' using the backpropagation algorithm.
Due to the high complexity of the architecture of these models, it is very hard, if not impossible, to understand what every single layer of the model is exactly doing.
This is why it is very difficult to debug these models, and why we can refer to complex \acp{ANN} as black-boxes.

\textbf{Vulnerability to attacks}.
\acp{ANN} and deep models consisting of many \acp{ANN} can be vulnerable to attacks.
These attacks consist in providing some input in the input space so that the network produces a wrong output.
For example, in the self-driving vehicle problem, a pattern or some black rectangles arranged in some way on a street sign, apparently harmless to the human eye, can disrupt the detection of the signal by the perception unit.
An attacker can put stickers on a road sign so that self-driving vehicles misread the signal, or ignore it altogether.
This can lead to catastrophic consequences in a road traffic scenario, and it is the reason why the vulnerability to this kind of attack should be considered when validating a self-driving vehicle.
a survey on currently known vulnerabilities to attacks of computer vision models based on deep learning can be found in Akhtar et al.~\cite{Akhtar2018}.

\section{Safety Requirements of Self-Driving Vehicles}

\begin{figure}[t]
    \centering
    \includegraphics[width=\linewidth]{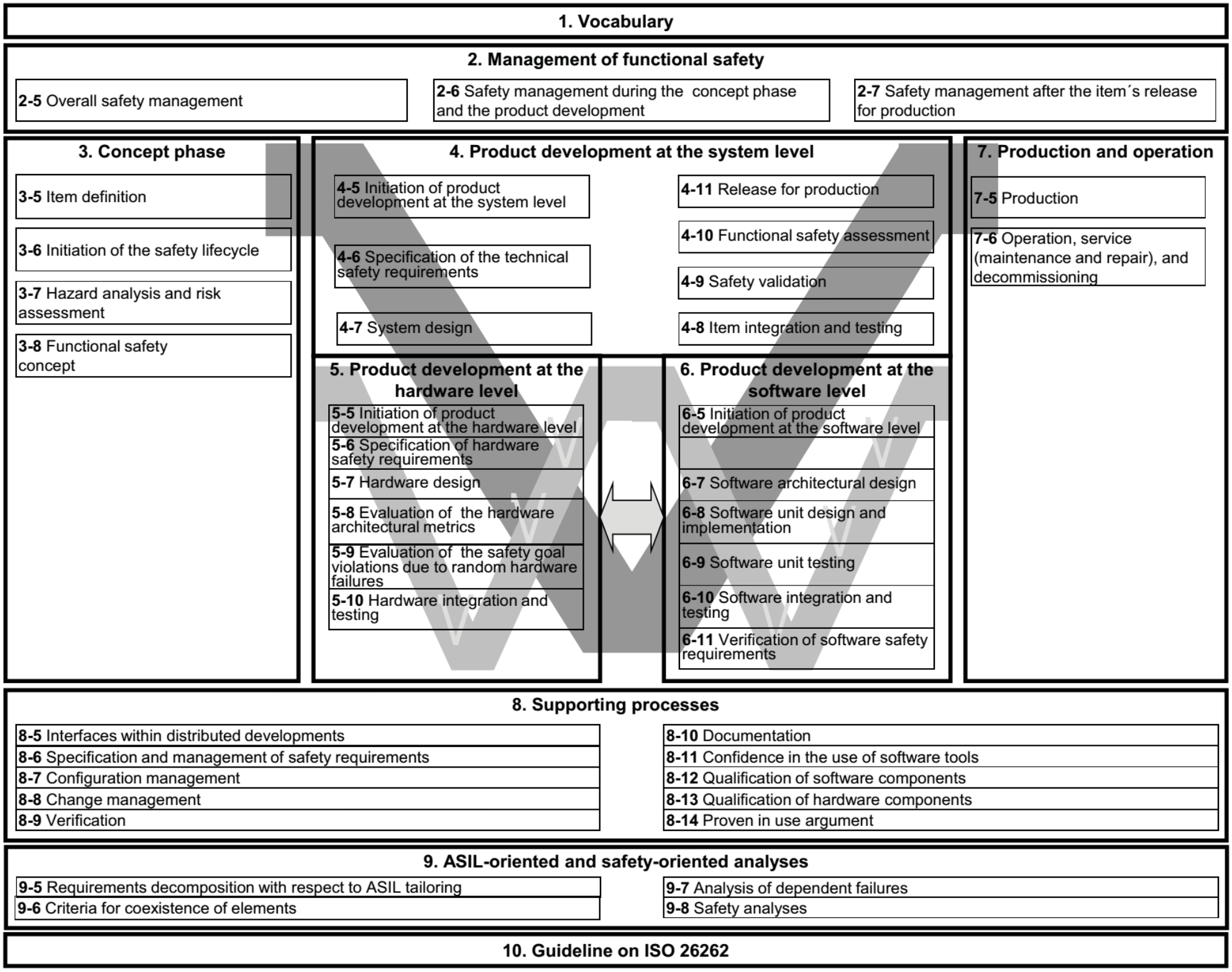}
    \caption{V-model of the ISO 26262~\cite{Jeon2011}. This model is a good starting point for defining a validation model for self-driving vehicles.}
    \label{fig:v_model}
\end{figure}

Self-driving vehicles are supposed to navigate through traffic, where a large number of risks are ever-present. Therefore, there is a need for a set of safety requirements that they should meet.
To the best of our knowledge, a standard that defines safety requirements for self-driving vehicles does not yet exist.
The ISO 26262~\cite{Jeon2011}, which is the standard that defines the safety requirements of electric and electronic systems of vehicles, is a good starting point for defining an ad-hoc standard.
ISO 26262 covers every phase of the development and production process using the V-model (Figure~\ref{fig:v_model})~\cite{Clark2009} as a reference validation model~\cite{Jeon2011}.

To measure the risk of failure of a specific component, ISO 26262 uses a risk classification scheme called \ac{ASIL}.
There are four \ac{ASIL} levels from A to D, A indicating the lowest risk and D indicating the highest.
The risk is determined by three factors: severity, which indicates the possible damage to passengers and property from the failure; exposure, which indicates the probability of a failure to occur; and controllability, which indicates how much the situation can be controlled when the failure occurs.
Once these factors are determined, the \ac{ASIL} level can be determined from a table that can be found in the ISO 26262~\cite{Jeon2011}.

The ISO has recently been developing a standard specifically for self-driving vehicles, named ISO/PAS 21448~\cite{schnellbach2019}.
It covers the cases in which a safety hazard can arise without it being caused by a system failure.
The difference with the ISO 26262 is that, instead of focusing on system failures, it focuses on malfunctions that can be caused by bad system design.
Both ISO 26262 and ISO/PAS 21448 should be taken into account when designing a self-driving vehicle, as they complement each other.
The limitation of ISO 26262 and ISO/PAS 21448, however, is that they are designed for self-driving vehicles supervised by a human.

Koopman et al.~\cite{koopman2019} also proposed a new standard for self-driving vehicles, named UL 4600.
This standard, instead of focusing on hardware and software specifications, remains technologically neutral, focusing instead on the safety principles.
The difference with the ISO 26262 and ISO/PAS 21448, is that the UL 4600 is specifically designed for vehicles with levels 4 or 5 of driving automation, where the driver is completely left out of the loop.
The limitation of this standard is that it does not provide benchmarks for testing on a road, or criteria for safety or performance.

\subsection{Driver out of the loop}

To validate vehicles with levels 4 or 5 of driving automation (Table~\ref{tab:SAE_EJ3016}), the driver should be left out of the loop during the testing phase.
This is because the vehicle should be able to recognize and deal with potentially catastrophic situations by itself, without the intervention of a human driver.
During the validation phase, it is useful to observe how a self-driving vehicle reacts to dangerous situations by itself, to discover in which kind of scenarios the vehicle needs more training.
This is the main reason why it is dangerous to validate a self-driving vehicle on a public road---it is unacceptable to release a vehicle which is a potential hazard.

\subsection{Accepted failure rate}

Failure rate is the frequency at which a catastrophic failure in a fleet of self-driving vehicles occurs~\cite{Koopman2016}.
In a hypothetical fleet of one million vehicles operated one hour per day, if the accepted failure rate is one catastrophic failure every 1000 days, Koopman et al.~\cite{Koopman2016} argue that the average time between catastrophic failures should be one billion hours.
This means that to validate such a failure rate, several billion hours of testing should be carried to be statistically significant, which is clearly unfeasible in the real world.

\subsection{Hazards}

Driving presents numerous hazards arising from nature and the behavior of other road users, such as bad weather, drivers that break rules either by distraction or intentionally, or animals suddenly crossing the street.
A fully autonomous vehicle should be able to identify these hazards, should drive carefully to avoid a potentially catastrophic scenario, and should know how to intervene on the controls if it finds itself in such a scenario.
Since many of these hazards occur randomly and cannot be predicted, the only practical way to validate the vehicle whenever they occur is to simulate them using a validation framework.

\subsection{Traffic laws}

A fully autonomous vehicle should observe traffic laws when driving on a public road.
Since traffic laws cannot be simply inferred from training alone, they should be manually hard-coded.
It is worth mentioning that traffic laws change over time and from country to country, which means that a self-driving vehicle should be regularly kept up to date, and should also be able to switch from one set of laws to another when it crosses a border between countries.

A self-driving vehicle should also be able to handle exceptions and conflicting road signs.
For example, if the vehicle sees a road marking that signals that overtaking is allowed (dashed line), but it also sees a vertical sign that forbids overtaking, the vehicle should know that the vertical sign has a priority on the road marking.
Hence, the vehicle should decide that it is not allowed to overtake in that particular road section.
This situation is very common during roadworks.

Since breaking traffic laws can cause potentially catastrophic scenarios, e.g. crossing a busy intersection with a red light, validating the lawful behavior of a self-driving vehicle should be done with a validation framework.
This also raises the question: who has the penal responsibility for a fatal crash caused by a self-driving vehicle because it broke a traffic law?
New laws and regulations written specifically for self-driving vehicles are needed.

\section{Validation Approaches and Techniques}\label{sec:approaches}

As discussed in the previous section, testing is impractical to do in the real world. We would need billions of hours of testing, in the traffic and with the driver out of the loop, where so many things can go wrong.
Therefore, simulations on validation frameworks seem to be the best solution at the state of the art for testing self-driving vehicles.
Validation frameworks are usually based on simulating frameworks.

Many frameworks exist for simulating the behavior of robots in a three-dimensional environment, e.g. Gazebo in conjunction with the ROS interface~\cite{Koenig2004}, or more specific simulators for self-driving vehicles.
Specific simulators include SUMO~\cite{SUMO2012}, an urban mobility simulator; Pro-SiVIC~\cite{Hiblot2010}, a simulation framework that can easily integrate sensors of self-driving vehicles; CarSim~\footnote{\url{https://www.carsim.com/}}, a vehicle and traffic simulator that accurately simulates physics; and SCANeR by AVSimulation~\footnote{\url{https://www.avsimulation.fr/solutions/}}, a simulation framework especially focused on making realistic scenarios.

\subsection{Simulating scenarios}\label{sec:scenarios}

A self-driving vehicle, once deployed, needs to drive through complex urban environments, where it can encounter signals, intersections, and traffic.
Simulating realistic scenarios is, therefore, an important part of testing self-driving vehicles.
These scenarios should represent real-world streets as accurately as possible, and should also simulate the traffic as realistically as possible.

\subsubsection{Closed world mapping}

Closed-world mapping is a useful method for simulating a realistic urban scenario, including all its physical features and its traffic laws~\cite{Zofka2016a}.
In the context of validating self-driving vehicles, a closed world model should include a 3D model of the closed world to be mapped, its physical laws, and its traffic laws.
The 3D model should include everything that is found in the real world, such as horizontal signs (e.g. lanes, parking signs), vertical signs (e.g. traffic lights, stop signs), temporary barriers in case of roadworks, random obstacles, etc.

Zofka et al.~\cite{Zofka2016a}, for the 2016 AUDI Autonomous Driving Cup, demonstrated how to simulate a closed world for testing self-driving vehicles.
Their framework is capable of simulating all the physical features of a realistic urban scenario, and to verify that the self-driving vehicles also follow the traffic laws.
Lattarulo et al.~\cite{Lattarulo2017} also created a framework that contains a simulator of a closed world, in an urban setting consisting of intersections and roundabouts.

\subsubsection{Simulating the traffic}

The scenarios that a vehicle can encounter in the traffic are virtually infinite.
There is a variety of roads, intersections, roundabouts with many different shapes, and vehicles of different shapes and sizes that navigate through them every day.
Plus, it is not to exclude that other drivers or self-driving vehicles can violate traffic laws, unintentionally or even on purpose.
A good framework should test self-driving vehicles with a comprehensive set of traffic scenarios so that the vehicle can be tested in as many traffic situations as possible.
simulating the traffic scenarios, therefore, is very time consuming and costly in terms of computational resources.

To cover every possible scenario, simulations of parts of scenarios can be optimally combined so that as many scenarios as possible are covered.
This can be done by splitting scenarios into multiple sub-tasks, where every sub-task is fulfilled by its own simulation.
Zofka et al.~\cite{Zofka2016} have shown that, with this approach, a framework can achieve a larger coverage of possible scenarios than simulating whole scenarios.
This way, testing scenarios can be done in a shorter time and with lower requirements of computational resources.

Traffic scenarios can also be represented by game-theoretic modeling.
Li et al.~\cite{Li2018} proposed a model based on the idea that different agents follow different levels of reasoning~\cite{Stahl1995, Costa2009}.
A \mbox{level-0} agent behaves without considering the actions of other agents in the system, a \mbox{level-1} agent assumes that other agents are \mbox{level-0} and behaves accordingly, a \mbox{level-2} agent assumes that other agents are \mbox{level-1} and so on.
As reported by Li et al.~\cite{Li2018}, this model is good because this multi-level reasoning has been observed in humans through multiple experiments~\cite{Hedden2002, Costa2006}.

\subsection{Image generation}\label{sec:image_generation}

Adding artificial rain, or fog, or distortions to the images perceived by self-driving vehicles, can have a dramatic effect on their behavior, as shown by Pei et al.~\cite{Pei2017}.
This was confirmed later by Tian et al.~\cite{Tian2018}, which have also shown that the same thing happens when changing the lightning conditions of such images. 
This implies that self-driving vehicles should be trained to be able to deal with all kinds of meteorological scenarios and possible lighting conditions.
This is practically unfeasible, and even thousands of hours of recording of driving in the real world cannot cover every possible scenario~\cite{leudet2019ailivesim}.
To increase the coverage of scenarios, both for training and testing purposes, real images can be artificially transformed into different scenarios.
For example, a recording of a bright sunny day can be transformed into an image of a foggy or a rainy day.
To transform images and still keep their realistic look, complex machine learning models based on deep learning can be used~\cite{Pei2017, Tian2018, Zhang2018}.

\acp{GAN}, a framework originally developed by Goodfellow et al.~\cite{Goodfellow2014}, has successfully been applied to the generation of artificial photo-realistic images~\cite{Denton2015, Zhang2017}.
A state-of-the-art approach by Zhang et al.~\cite{Zhang2018} successfully applied \acp{GAN} for generating high-quality photo-realistic images for training self-driving vehicles.
This includes, for example, transforming the image of a road taken in summer into a realistic-looking winter snowy road.
The generation of such images allows the generation of comprehensive training and test datasets in every possible meteorological and lighting condition, which is important for validating self-driving vehicles.

\subsection{V2V and V2I networks}\label{sec:v2v_v2i_networks}

\ac{V2V} and \ac{V2I} communication architectures, generalized as V2X, can be used to tackle some problems related to traffic.
As we have discussed in the introduction, this is especially important in smart cities, where self-driving vehicles have the potential to improve the traffic flow.
V2X has been proposed for numerous applications, such as intelligent traffic management at intersections~\cite{Bento2012}, cooperative driving systems~\cite{Jia2016}, and vehicular density estimation systems~\cite{Barrachina2013}.
This, however, presents numerous challenges: the timing of the messages is affected by the surrounding environment; the wireless connection between moving vehicles is by its nature unstable; the connection might be vulnerable to attacks.

To increase the safety and efficiency of self-driving vehicles, different V2X architectures have been proposed~\cite{Harding2014}.
To provide a simulation framework for these challenges, Zheng et al.~\cite{Zheng2016} proposed a cross-layer modeling, exploration and validation framework called CONVINCE.
It can be used to simulate how external situations can impact the performance of V2X communications.
Specifically, in their study, they show a case study on how a flooding attack can cause packet loss and cause problems in V2X communications.
Another framework based on Dynacar, developed by Lattarulo et al.~\cite{Lattarulo2017}, contains a module that enables the testing of V2X communications.

\subsection{Fault injection}\label{sec:fault_injection}

Fault injection is a technique that consists in stressing a system in an unusual way, to check how it reacts.
For example, if we wish to test the hardware of a system we can subject it to different temperatures or different voltages.
If we wish to test the software, we can feed it unusual inputs.
Usually, fault injection on software is carried in a way that maximizes the coverage of code paths.

Self-driving vehicles can have many kinds of faults, both of hardware or software nature.
For example, a sensor or one of the cameras can fail, or the car can react in the wrong way to a scenario never encountered before.
Another possibility is that the perception unit is disrupted by a pattern placed on a traffic sign by an attacker.
Validating self-driving vehicles can be done, for example, by injecting its sensors with distorted inputs~\cite{Jha2018, Rubaiyat2018, Pei2017}.
Faults and errors could also be injected into the self-driving system itself~\cite{Jha2018, Jha2019}.

Jha et al.~\cite{Jha2019a} proposed a clever approach to perform fault injection on self-driving vehicles, named ``Bayesian fault injection''.
It consists in using machine learning to automatically mine situations and faults that can critically affect a self-driving vehicle.

\section{Roadmap}\label{sec:roadmap}

Let us now conclude the chapter, discussing what the ideal framework for the validation of self-driving vehicles should be, and a possible roadmap for solving the open problems.

\subsection{State-of-the-art}

The main purpose of validation frameworks for self-driving vehicles should be testing them in every possible urban and extra-urban scenario.
Each presents its own challenges: the car should be able to navigate through the traffic in urban settings, and should also be able to drive fast on the highway while maintaining a high level of safety by monitoring the surrounding environment.

The best approaches at the state of the art for the mapping of a closed world can be found in Zofka et al.~\cite{Zofka2016a} and Lattarulo et al.~\cite{Lattarulo2017}.
Specifically to the generation of traffic scenarios, the approaches found in Li et al.~\cite{Li2018} and Zofka et al.~\cite{Zofka2016} should be considered, namely modeling the traffic using a game-theoretic approach, but also optimally considering different scenarios to optimize the coverage of all possible scenarios.
It could be useful to equip every self-driving vehicle with a black box and use the data recorded during crashes to infer which are the real world situations most likely to lead to an accident.

Since the sensors of a self-driving vehicle include also cameras, they also need to be tested with images as photo-realistic as possible.
The best approach at the current state of the art seems to be the generation of images through \acp{GAN}~\cite{Zhang2018}.

V2X should be an important component of self-driving vehicles.
As we discussed, though, it can be subject to problems and attacks.
At the current state of the art, the best validation approaches can be found in Zheng et al.~\cite{Zheng2016} and Lattarulo et al.~\cite{Lattarulo2017}.

The software of self-driving vehicles should be thoroughly tested for faults and errors, and its behavior when receiving distorted inputs should also be thoroughly tested.
As we discussed, there is a framework at the state of the art that combines these two important issues, that can be found in Jha et al.~\cite{Jha2019a}.

Once issues in a self-driving vehicle are spotted by the validation framework, they need to be addressed by making changes to the hardware or software of the vehicle.
Whenever a change is made to the hardware or software that governs a self-driving vehicle, the vehicle should be validated again.
This raises the question, should the whole validation process be repeated, or should only a subset of the whole validation process be carried?
If the second case is possible, an ideal validation framework should be able to select smartly the subset of tests that should be carried out after a change in one of the components.

\subsection{Future work}

The validation of self-driving vehicles is still an open research problem.
Let us discuss some issues that future research could focus on.

\subsubsection{Safety requirements}

Currently, no special requirements exist for periodical inspections of self-driving, they have to pass the same criteria that normal vehicles have to.
A well-defined set of safety requirements for self-driving vehicles can also benefit the development of validation frameworks that will be needed for this specific technical inspection.

\subsubsection{V2X}

Communication between self-driving vehicles increases their safety and efficiency.
To the best of our knowledge, no standard exists on V2V and V2I communications between self-driving vehicles.
A standard should be developed so that different manufacturers can develop vehicles able to communicate with each other.
In turn, this will also allow easier development of validation frameworks for self-driving vehicles, since frameworks will not need to have different communication modules for different manufacturers.

\subsubsection{Vulnerabilities}

Only a few of the validation frameworks at the state of the art focus on possible vulnerabilities of self-driving vehicles~\cite{Zheng2016, Pei2017}.
Future validation frameworks can benefit from more thorough testing of vulnerabilities, based on recent works on sensor and model vulnerability~\cite{Amoozadeh2015, Petit2015, Jafarnejad2015, Lim2018, Akhtar2018}.

\bibliographystyle{abbrv}  
\bibliography{references}

\end{document}